# A car-following model with behavioural adaptation to road geometry

Fabrizio Pelella[a], Gaetano Zaccaria[a], Vincenzo Punzo[a], Marcello Montanino[a,*]

[a] Department of Civil, Environmental and Architectural Engineering, University of Naples Federico II, Via Claudio 21, 80125, Napoli (NA), Italy

* Corresponding author: Ph.: +39 081 7683770

**Email addresses:**
fabrizio.pelella@unina.it
gaetano.zaccaria@unina.it
vincenzo.punzo@unina.it
marcello.montanino@unina.it

Pelella, Zaccaria, Punzo, Montanino

**ABSTRACT**

Understanding the effect of road geometry on human driving behaviour is essential for both road safety studies and traffic microsimulation. Research on this topic is still limited, mainly focusing on free-flow traffic and not adequately considering the influence of curvature on car-following dynamics. This work attempts to investigate this issue and model the adaptation of car-following behaviour to horizontal curvature. For this purpose, the maximum desired speed – which mainly determines the free-flow dynamics – is expressed as a parsimonious function of the curvature. A spatial anticipation mechanism is also included in order to realistically describe the driving behaviour when approaching or exiting from curves. The accuracy of the augmented model is evaluated using the Modified Intelligent Driver Model (M-IDM) and trajectory data from free-flow and car-following traffic (Naples data and Zen Traffic Data). The results show that a significant improvement is achieved in free-flow dynamics. In car-following situations, improvements are mainly observed at high speed and are dependent on the observed driver. Overall, the analysis highlights the lack of sufficiently spatially extended trajectory data to calibrate and evaluate such driving behaviours.

## 1. INTRODUCTION

While driving, the driver adapts his behaviour to stimuli from the traffic and road geometry. In free-flow conditions, the driver's response to stimuli from the road infrastructure becomes primary. A significant element is certainly the horizontal curvature, which affects the vehicle dynamics. In fact, driver comfort is influenced by the centripetal force acting on the vehicle (a function of vehicle speed and road curvature). For this reason, speed adaptation takes place to restore an acceptable level of comfort when the centripetal force increases. Such adaptation may be predominant in the driver's response to car-following stimuli.

Although the relationship between speed and horizontal road geometry has been thoroughly investigated in road design and road safety studies (e.g., Oviedo-Trespalacios et al., 2017; Abu Addous, 2021), it is largely overlooked in traffic flow theory. Even commercial software, such as Aimsun, neglects the driver's response to bends and its impact on speed.

Among the available studies in the field of traffic flow theory, most of them have focused on free-flow dynamics (e.g., Li and Smith, 2014; Cheng et al., 2019). Very few studies have attempted to model the effect of curvature on driver's response in car-following situations (Partouche et al., 2007; Chan et al., 2007; Zhu and Zhang, 2012; Yoshizawa et al., 2012; Li and Smith., 2014; Zhu and Yu, 2014; Jin et al., 2017; Zheng et al., 2017; He et al., 2022; Han et al. 2024; Yang et al., 2024).

A review of these studies revealed some limitations. First, a quantitative analysis of the effect of speed adaptation to road geometry on model accuracy is only provided in free-flow traffic (Li and Smith, 2014; Cheng et al., 2019). In car-following, most studies have evaluated the impact of such adaptation on the linear string stability behaviour of the model (Zhu et al., 2012; Zhu et al., 2014; Jin et al., 2017; Zheng et al., 2017; Han et al., 2024). To the best of our knowledge, no quantitative analyses of model accuracy by calibration against observed trajectory data have been performed in the literature.

Second, the proposed approaches have been shown to be unable to model the driver's spatial anticipation when approaching or exiting curves (Partouche et al., 2007; Chan et al., 2007; Yoshizawa et al., 2012), showing a systematic lag in simulated speed profiles compared to observed data. This fact is clearly illustrated in He et al. (2022), which highlights the need for more appropriate model formulations.

This study aims to overcome these two limitations. A parsimonious model of the adaptation of car-following behaviour to horizontal curvature is formulated. In particular, the maximum desired speed model parameter is expressed as a function of curvature. A spatial anticipation mechanism is also included to realistically describe the driving behaviour when approaching or exiting from curves. The accuracy of the proposed model extension is evaluated using the Modified Intelligent Driver Model (*M-IDM*) and trajectory data from free-flow and car-following traffic (Naples data and Zen Traffic Data).

The paper is structured as follows. Section 2 presents the model formulation. Section 3 describes the trajectory data used for model calibration. The results of the calibration experiment are presented in section 4. A conclusion concludes the paper.

## 2. INCORPORATING SPEED ADAPTATION TO HORIZONTAL CURVES IN CAR-FOLLOWING MODELS

The speed adaptation to horizontal curvature is modelled by expressing the maximum desired speed parameter $v_0$ of car-following models as a function of the perceived curve radius $R_p(x)$ at longitudinal position $x$. In particular:

$$v_0(t) = v_{0,straight} - \frac{\gamma}{R_p(x)} \qquad (1)$$

where $v_{0,straight}$ is the maximum desired speed of the driver on straights, and $\gamma$ is a curvature factor, which describes the behavioural effect that horizontal curvature has on driver's desired speed. The perceived curve radius allows to take into account human driver spatial anticipation behaviours. In fact, the driver anticipates a downstream variation of horizontal curvature by either decelerating prior to approaching a curve or accelerating before exiting a curve. Therefore, the driver reacts to a perceived radius $R_p(x)$ at location $x$ which may differ from the actual radius $R(x)$, depending on the level of spatial anticipation. Such anticipation is modelled be means of time-varying look-ahead distance, as a function of speed:

$$R_p(x) = \begin{cases} R(x(t) + T_{ant} \cdot v(t)) & if\ R(x(t) + T_{ant} \cdot v(t)) \leq R_{lim} \\ \infty \end{cases} \quad (2)$$

where $T_{ant}$ is a constant load-ahead time-headway. In addition, Eq. 2 also considers a human perception threshold of horizontal curvature, through an indifference threshold $R_{lim}$.

In this work, the proposed modelling framework is applied to a modified version of the Intelligent Driver Model (M-IDM), which is derived by combining the original model formulation by Treiber et al. (2000) and a critical regime behaviour introduced in Tian et al. (2016).[1]

$$a(t) = \begin{cases} a \cdot \left[1 - \left(\frac{v(t)}{v_0(t)}\right)^\delta - \left(\frac{s^*(t)}{s(t)}\right)^2\right] & if\ s^*(t) \leq s(t) \\ \begin{cases} a \cdot \left[1 - \left(\frac{s^*}{s(t)}\right)^2\right] & if\ v(t) \leq v_{crit} \\ min\left\{a \cdot \left[1 - \left(\frac{s^*}{s(t)}\right)^2\right], -b\right\} \end{cases} \end{cases} \quad (3)$$

where $v(t)$ and $s(t)$ are the ego-vehicle speed and the net inter-vehicle spacing between the leader and the ego vehicle, $a$ and $b$ are the maximum acceleration and normal deceleration rate of the ego-vehicle, $\delta$ is a model parameter, $v_{crit}$ is the critical speed identifying the critical regime behaviour, $s_0$ is the minimum net inter-vehicle spacing at stop. $s^*(t)$ is the desired net spacing, defined as $s^* = s_0 + \max\left(0, v(t) \cdot T - \frac{v(t) \cdot \Delta v(t)}{2\sqrt{ab}}\right)$, where $T$ is the desired time-headway, and $\Delta v(t)$ is the leader-follower speed differential. A ballistic integration scheme $\Delta t = 0.1\ s$ is adopted to update ego-vehicle speed and position (Punzo et al., 2021).

In the following, we will refer to the *M-IDM-r* as the *augmented model with speed adaption* due to horizontal curves modelled according to Eq. 1, and to the *M-IDM* as the model without speed adaption, i.e., $\gamma = 0$ in Eq. 1.

## 3. EXPERIMENTAL DATA

In the current work, three vehicle trajectory datasets have been used to evaluate the accuracy of the augmented model: *i)* the *Asse Mediano* dataset which contains free-flow trajectory data of an isolated vehicle collected in Naples through a GPS and OBD reader, *ii)* a selection of vehicle trajectories from the *Zen Traffic Data* project, including car-following dynamics at both high and low speeds, and *iii)* the vehicle trajectories collected on *Variante Solfatara* in the car-following experiment conducted in Naples in early 2000 (Punzo and Simonelli, 2005).

Concerning the *Asse Mediano* dataset, the experiments have been carried out on the SS 162 NC "Asse Mediano" in June 2024 (see Figure 1). It is a rural road with two lanes per direction. The study section develops on a viaduct and stretches for approximately 12 km, from the "A1" national motorway to the "SP 335 – Ponti della Valle". The road is mostly flat, with very low gradients and a difference in height between the end points

---

[1] The proposed model differs from the one in Tian et al. (2016), since for $s^*(t) \leq s(t)$, the original IDM model formulation is adopted. In Tian et al. (2016), instead, for $s^*(t) \leq s(t)$, $a(t) = a \cdot \left[1 - \left(\frac{v(t)}{v_0(t)}\right)^\delta\right] \cdot \left[1 - \left(\frac{s^*(t)}{s(t)}\right)^2\right]$.

Pelella, Zaccaria, Punzo, Montanino

of 74 m, while the radii vary from 188 m and 999 m, showing a high degree of tortuosity, alternating straights and curves with a small radius. The experiment was carried out on 26 June 2024, from 5:30 AM to 7:00 AM, on Sunday in a sunny day, in free-flow conditions. Using a GPS and OBD reader, detailed vehicle dynamics and geographical data were collected at 10 Hz. A total of eight vehicle trajectories were collected.

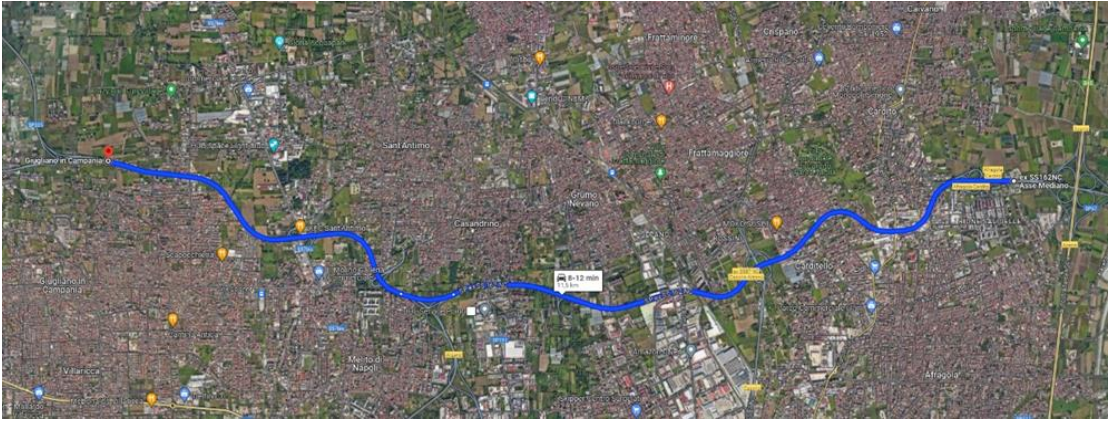

**Figure 1: SS162NC "Asse Mediano" experimental section.**

In the second database (*ZEN*), two subsets of vehicle trajectory data at 10 Hz were extracted from the Route 4 dataset, the only experimental location showing horizontal curves. The section (Figure 2), with a total length of 1.6 km, has two straights and one curve of 500 m radius. Two subsets of car-following trajectory data were extracted: one where the maximum observed ego-vehicle speed is above 90 kph (high speed data, 33 trajectories), and one where it is lower than 60 kph (low speed data, 36 trajectories). In both datasets, the ego-vehicle is always a car (motorcycles and trucks were excluded), and no lane-changing occur (neither of the ego, or the leading vehicle). All trajectories have a minimum duration of 50 seconds.

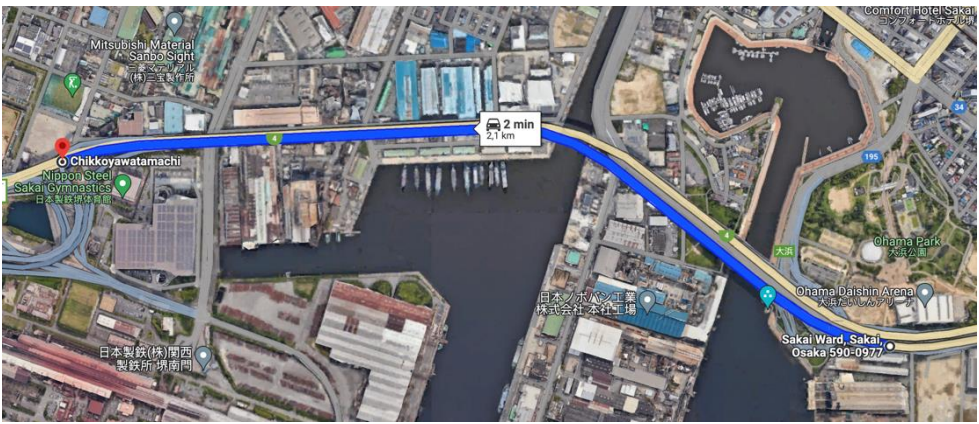

**Figure 2: Hanshin Expressway Route 4 experimental section.**

The third dataset (*Variante Solfatara*) consists of car-following vehicle trajectory data at 10 Hz collected on "Variante Solfatara" in October 2002 with a platoon of 4 vehicles along a rural road, with single carriageway and one lane per direction (see Punzo and Simonelli 2005 for a more detailed description). The study section (Figure 3) has 7 curves with radii varying from 60 m to 999 m.

Pelella, Zaccaria, Punzo, Montanino

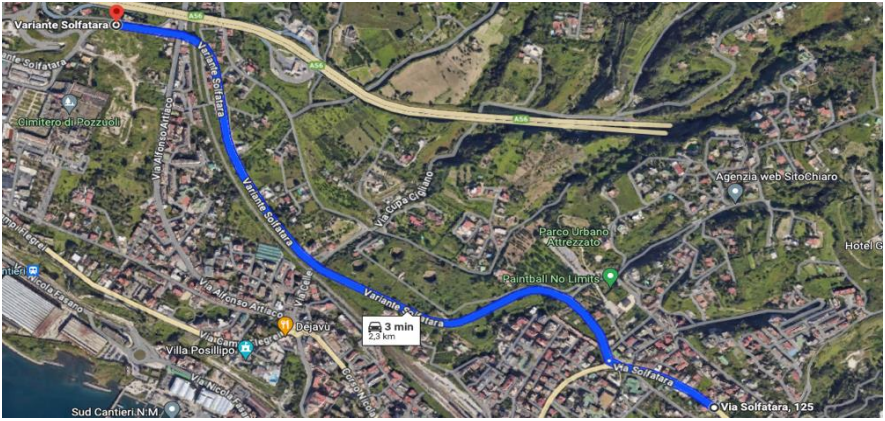

**Figure 3: Variante Solfatara experimental section.**

## 4. MODEL CALIBRATION

### 4.1. Methodology

A generalized Least Square Estimation method is applied to calibrate model parameters against observed ego-vehicle dynamics. The general formulation of the calibration problem is taken from Punzo et al. (2021). In particular, Table 1 shows the lower and the upper bounds of model parameters adopted in calibration.

**Table 1: Model parameter bounds**

| Parameter | Lower bound | Upper bound |
|---|---|---|
| a | 0.1 | 5 |
| b | 0.1 | 5 |
| T | 0.1 | 4 |
| $\delta$ | 0 | 10 |
| $v_{0,straight}$ | $\max\{0, \max(v_{obs}) - 10\}$ | $\max(v_{obs}) + 10$ |
| $v_{crit}$ | 0 | $\max(v_{obs}) + 10$ |
| $s_0$ | 0.1 | 10 |
| $\gamma$ | -1e4 | 0 |
| $T_{ant}$ | 0.1 | 4 |
| $R_{lim}$ | 0 | 1e6 |

Based on the guideline provided in Punzo et al. (2021), the Normalized Root Mean Squared Error of speed and spacing, *NRMSE(s,v)*, is used as GOF in the optimization problem. The Genetic Algorithm coded in Matlab is used to solve the optimization problem. The following setting for hyper-parameters is adopted: population size equal to 100; number of generation equal to 10,000; a tolerance value equal to 1e$^{-4}$ and limit for stall generations with no improvements on the objective function equal to 100.

Both the *M-IDM* and the *M-IDM-r* were calibrated against experimental data. Results are discussed in the next section.

## 4.2. Results

The boxplot of model calibration errors of *M-IDM* and *M-IDM-r* on vehicle trajectory data from the four datasets are reported in Figure 3. Figure 4 shows, instead, the boxplot of the relative error variation – computed on each trajectory $i$ – between the M-IDM-r and the M-IDM, i.e.:

$$\Delta NRMSE(s,v)_i = \frac{NRMSE(s,v)_i^{m-idm-r} - NRMSE(s,v)_i^{m-idm}}{NRMSE(s,v)_i^{m-idm}}$$

In the figures, markers identify error values falling in the *first* and *last quartile* of the error distribution across different vehicle trajectories.

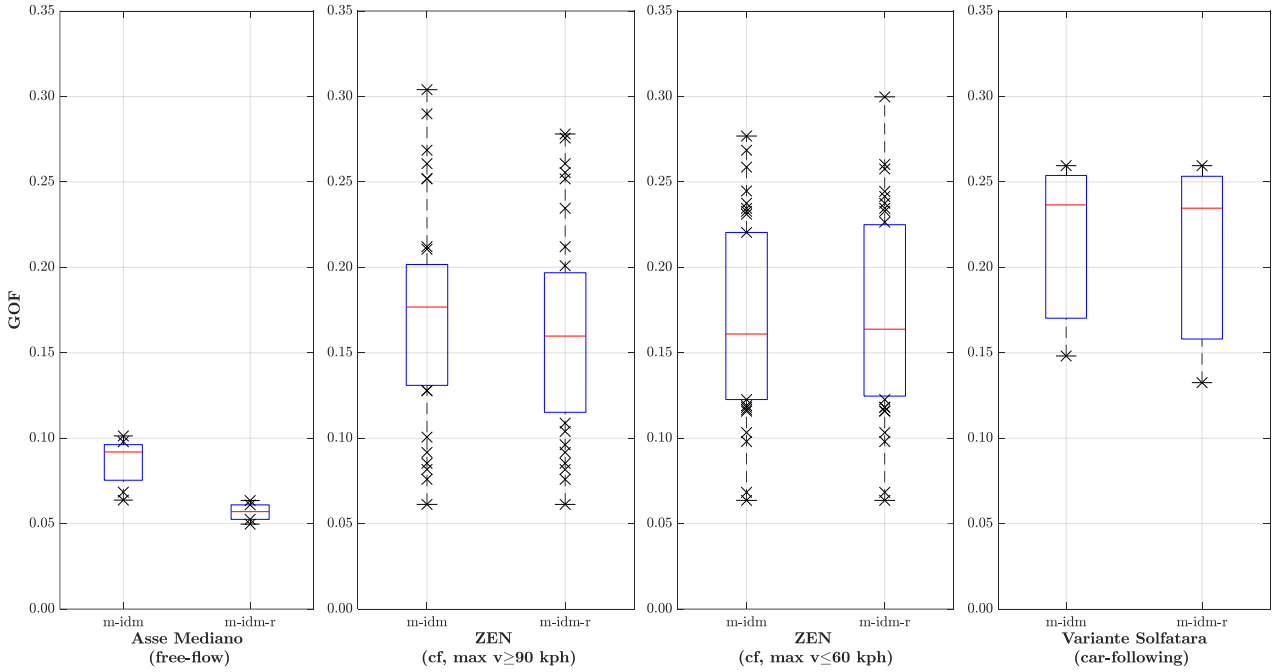

**Figure 4: Boxplots of the three datasets, representing the absolute error (*NRMSE*) of trajectories.**

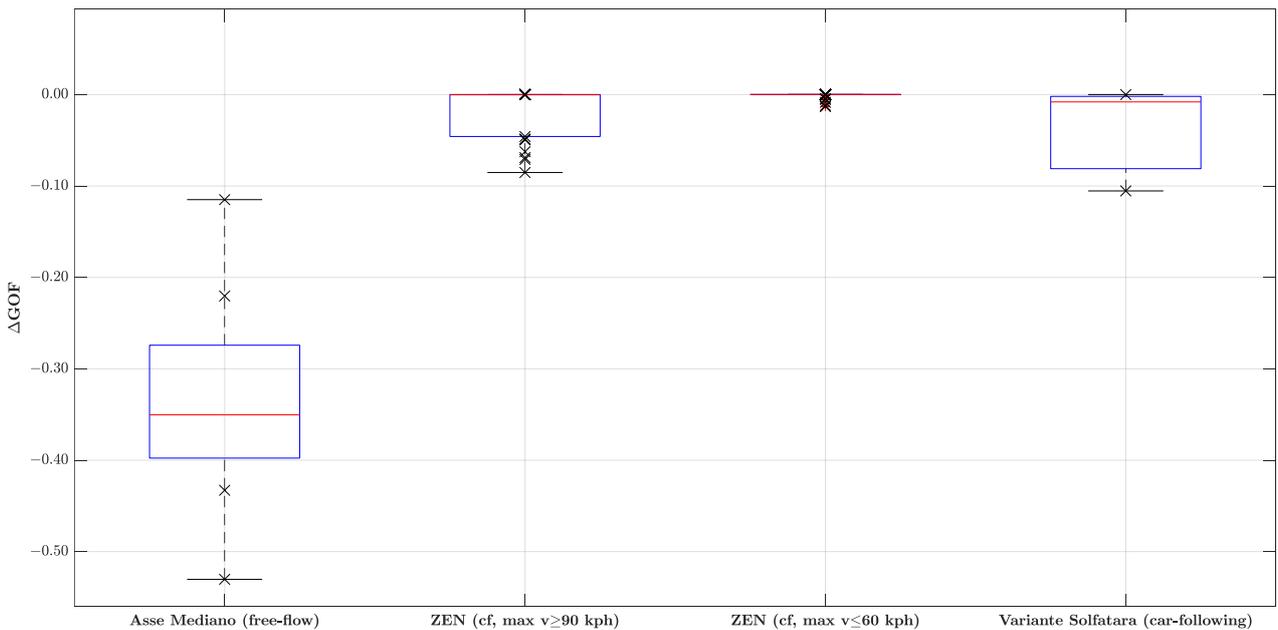

**Figure 5: Boxplots of the three datasets, representing the relative error (*NRMSE*) of trajectories.**

Results on free-flow data (see *Asse Mediano* boxplots figures), clearly show that incorporating speed adaptation improves model accuracy of at least 12% on all trajectories (35% on average). In car-following situations, improvements are less intense, and mostly occur at high speeds (see ZEN dataset with maximum speed above 90 kph, and *Variante Solfatara* data). In fact, at low speeds, the speed adaptation to car-following stimuli is prevailing.

Focusing on the Zen data with maximum speed above 90 kph, boxplots show that in 25% of trajectories (8 out of 33 trajectories) an error reduction of at least 5% occurred, while in *Variante Solfatara* dataset, the augmented model achieved an error reduction up to 10%.

Overall, if these results are promising, they also highlight the lack of trajectory data sufficiently extended in space to calibrate and evaluate such adaptive driving behaviour (please note, for instance, that in the ZEN datasets there is only one curve, while the *Variante Solfatara* dataset contains only three trajectories).

## 5. CONCLUSION

Modelling driver adaptation to horizontal curvature in car-following situations is a much-overlooked research topic in traffic flow theory. However, experience shows that such behaviour is a key component of human driving on curved roads. Of the few modelling approaches available, they have either been tested only in free-flow situations or have shown significant limitations for application in car-following situations (e.g. lack of spatial anticipation).

This study aims to overcome these limitations by developing a parsimonious model of car-following driving behaviour adaptation to horizontal curvature. In particular, the maximum desired speed model parameter is expressed as a function of the curvature. A mechanism of spatial anticipation is also included to realistically describe the human driving behaviour when approaching or exiting from road bends. The accuracy of the proposed model extension is evaluated using the Modified Intelligent Driver Model (M-IDM), and trajectory data from free-flow and car-following traffic (Naples data and Zen Traffic Data).

Results on free-flow data clearly show that incorporating speed adaptation significantly improves model accuracy. In car-following situations, the improvements are less pronounced and mostly occur at high speeds (in fact, speed adaptation to car-following stimuli dominates at low speeds).

This study also highlights that the quality and completeness of vehicle trajectory data – sufficiently extended to cover multiple curves of different radii – is key to effectively calibrate and validate such driving behaviours. If the preliminary results discussed here are promising, an extended car-following trajectory data collection campaign will be required to provide more comprehensive results on the accuracy of the augmented car-following model.

## ACKNOWLEDGMENTS

Research in this paper has been partially funded by the European Commission Horizon Europe project "i4Driving" (Grant Agreement ID 101076165).